\documentclass[journal]{IEEEtran}
\usepackage{amsmath,amsfonts}
\usepackage{algorithmic}
\usepackage{algorithm}
\usepackage{array}
\usepackage[caption=false]{subfig}
\usepackage{textcomp}
\usepackage{stfloats}
\usepackage{url}
\usepackage{verbatim}
\usepackage{graphicx}
\usepackage{cite}
\usepackage{threeparttable}
\usepackage{multirow}
\usepackage{booktabs}
\usepackage{xspace}
\usepackage{pifont}
\usepackage{color}
\usepackage{enumitem}
\usepackage{marvosym}
\usepackage{fontawesome}
\usepackage{balance}

\newcommand {\orthrus}{Orthrus\xspace}
\newcommand {\nodlink}{Nodlink\xspace}
\newcommand {\magic}{Magic\xspace}
\newcommand {\flash}{Flash\xspace}
\newcommand {\kairos}{Kairos\xspace}

\begin{document}

\title{How Far Should We Need to Go : Evaluate Provenance-based Intrusion Detection Systems in Industrial Scenarios}

\author{
\IEEEauthorblockN{
    Yue Xiao\IEEEauthorrefmark{2},
    Ling Jiang\IEEEauthorrefmark{3},
    Sen Nie\IEEEauthorrefmark{3}, 
    Ding Li\IEEEauthorrefmark{4},
    Shi Wu\IEEEauthorrefmark{3},
    Ke Xu\IEEEauthorrefmark{2}, and
    Qi Li\IEEEauthorrefmark{2}\textsuperscript{\Letter} \\
}
\IEEEauthorblockA{
    \IEEEauthorrefmark{2}Tsinghua University,
    \IEEEauthorrefmark{3}Tencent Security Keen Lab, 
    \IEEEauthorrefmark{4}Peking University, \\
    % 后面跟邮箱
    \IEEEauthorrefmark{2}xiaoy25@mails.tsinghua.edu.cn,
    \IEEEauthorrefmark{3}\{leviljiang, snie, shiwu\}@tencent.com,  \\
    \IEEEauthorrefmark{4}ding\_li@pku.edu.cn,
    \IEEEauthorrefmark{2}\{xuke, qli01\}@tsinghua.edu.cn
}
\thanks{
\textsuperscript{\Letter} is the corresponding author.
}
}



\maketitle

% \item We explain why PIDSes perform well or poorly from both qualitative and quantitative perspectives. In a word, we find that the primary factor affecting the performance is the feature learning task of a PIDS. The quantitative analysis shows that the Pearson correlation coefficient between the feature learning task and the AUC is approximately 0.73, with a p-value of 0.02.
%     \item We analyze the causes of high false positive rates and propose a mitigation method. We category the causes into three types: sparse activities, unknown activities and semantic changes of activities. Our proposed method can reduce the false positive rate from about 25\% to about 10\%, and reduce the manual effort by 2/3.
%     \item We find some problems may be overlooked but are important in practice. For example, because of the inconsistency between the target of fine-grained detection and that of investigation, high detection performance may also result in low investigation efficiency, exacerbating the alert fatigue problem.

\begin{abstract}
Provenance-based Intrusion Detection Systems (PIDSes) have been widely used to detect Advanced Persistent Threats (APTs). Although many studies achieve high performance in the evaluations of their original papers, their performance in industrial scenarios remains unclear. 
To fill this gap, we conduct the first systematic evaluation and analysis of PIDSes in industrial scenarios. 
We first analyze the differences between the data from DARPA datasets and that collected in industrial scenarios, identifying three main new characteristics in industry: heterogeneous multi-source inputs, more powerful attackers, and increasing benign activity complexity. We then build several datasets to evaluate five state-of-the-art PIDSes. The evaluation results reveal challenges for existing PIDSes, including poor portability across different hosts and platforms, low detection performance against real-world attacks, and high false positive rates with ever-changing benign activities. 
Based on the evaluation results and our industrial practices, we provide several insights to solve or explain the above problems. For example, we propose a method to mitigate the high false positives, which reduces manual effort by 2/3.
Finally, we propose several research suggestions to improve PIDSes.
\end{abstract}

\begin{IEEEkeywords}
Provenance-based Intrusion Detection System, Anomaly Detection
\end{IEEEkeywords}

\section{Introduction}
\label{sec:intro}

Cyber attacks increase rapidly in recent years. One type of attack is the Advanced Persistent Threat (APT), which uses a complex procedure to carry out attacks. APTs have raised serious concerns~\cite{apt-survey}.
To defend against APTs, Provenance-based Intrusion Detection Systems (PIDSes) are proposed.
PIDSes collect system logs (e.g., system calls in Linux), build provenance graphs, learn features of benign activities and identify the activities whose features deviate from benign ones.
PIDSes can effectively utilize the contextual information within the logs, making them popular in attack detection.

The main goal of PIDSes is to detect malicious entities (i.e., processes, files, sockets) in provenance graphs that contain a large number of benign entities.
To achieve this goal, many studies~\cite{SLEUTH, SAQL, MORSE, HOLMES, Poirot, RapSheet, p-gaussian, CAPTAIN, deeplog, UNICORN, SIGL, StreamSpot, Log2vec, provtalk, PROVDETECTOR, SHADEWATCHER, THREATRACE, PROGRAPHER, NODLINK, DISTDET, KAIROS, R-CAID, FLASH, MAGIC, orthrus,slot, STGAN, llm-detection} concentrate on designing a powerful PIDS. The most commonly used datasets for evaluating PIDSes are collected from DARPA Transparent Computing (TC) program, including DARPA-E3, DARPA-E5~\cite{dataset-DARPA}, and DARPA-OpTC~\cite{dataset-DARPAOpTC} datasets.
Although these PIDSes achieve high performance in evaluations of their original papers, their performance in practice remains unclear.

To fill this gap, we collect data from the users of two deployed services in a leading technology company. These two services include a security management service (such as~\cite{security-management-standards}) and a cloud workload service (such as~\cite{cloud-workload-protection-survey}). 
Our target is to evaluate the performance of PIDSes, explore the influence factors of the performance, provide insights based on the evaluation results and our industrial practices, and give suggestions for improving PIDSes.

We first analyze the differences between the data from DARPA datasets and that collected by us. We find three main characteristics of our data. (1) Heterogeneous multi-source inputs. The DARPA datasets concentrate on one host and one OS (i.e., Linux). In contrast, our data are collected from multiple services, including different operating systems (e.g., Linux and Windows) and different hosts. (2) More powerful attackers. The DARPA datasets are published in 2018 (E3) and 2020 (E5 and OpTC). Over time, attackers may use more advanced techniques to launch attacks. (3) Increasing benign activity complexity. The DARPA datasets are collected in a red-blue competition and the victims' activities are simple. In contrast, real users may exhibit various activities.  Detailed analyses are presented in \S~\ref{sec:industrial-character}.
These new characteristics may challenge the performance of PIDSes.

We build several datasets to examine the influences of the above characteristics. We select five state-of-the-art (SOTA) PIDSes for evaluation. These PIDSes cover different detection granularities~\cite{survey-eval} and utilize different detection techniques, representing the main types of current PIDSes. Our evaluation results challenge the existing PIDSes, summarized as follows:
\begin{itemize}[nosep]
% \begin{itemize}[nosep, leftmargin=*]
    % AUC decrease: 0.2013, 0.6372, 0.1326, 0.3917, -0.0244, average: 0.2677, from -0.0244 to 0.6372
    % AUC decrease: 0.7379 - 0.6645 = 0.0734, 0.9004 - 0.3116 = 0.5888, 0.9677 - 0.7219 = 0.2458, 0.9437 - 0.2310 = 0.7127, 0.7837 - 0.5027 = 0.2810, average: 0.3803, from 0.0734 to 0.7127
    \item Poor portability across different hosts and platforms. We find that the PIDSes trained on one host or platform do not perform well on others. The average AUC drops by 26.77\% when testing on different hosts and by 38.03\% when testing on different platforms.
    % AUCs (mining): 0.3943, 0.5375, 0.6419, 0.5577, 0.9217, average: 0.6106, from 0.3943 to 0.9217
    % AUCs (stealing): 0.4216, 0.4855, 0.5399, 0.4840, 0.9372, average: 0.5736, from 0.4216 to 0.9372
    \item Low detection performance against real-world attacks. We use a mining attack and a information stealing attack that happened in real world to evaluate. The AUCs of the PIDSes are from 39.43\% to 92.17\% (61.06\% on average), and from 42.16\% to 93.72\% (57.36\% on average), respectively.
    \item High false positive rates with ever-changing benign activities. When benign activities change frequently, three PIDSes suffer from high false positive rates (over 23\%), even if there are no attacks.
    % real-case-2': {'flash': {'train_time': 33.65, 'test_time': 4.33}, 'orthus': {'train_time': 55.95, 'test_time': 789.04}, 'magic': {'train_time': 98.22, 'test_time': 363.06}, 'karios': {'train_time': 3638.21, 'test_time': 333.06}, 'nodlink': {'train_time': 649.43, 'test_time': 239.41}}, 'real-case-stolen': {'orthus': {'train_time': 11.38, 'test_time': 60.27}, 'magic': {'train_time': 11.82, 'test_time': 4.26}, 'flash': {'train_time': 4.67, 'test_time': 0.41}, 'nodlink': {'train_time': 28.43, 'test_time': 13.75}, 'karios': {'train_time': 340.98, 'test_time': 31.38}}}
    % \item Significant time overhead for practical deployment. The time overhead of most PIDSes is unacceptable. With a graph containing about 10$^6$ nodes and edges, the training time and testing time is from 33.65 seconds to 3638.21 seconds and from 0.41 seconds to 789.04 seconds, respectively.
\end{itemize}

% 此外，我们基于实验结果，提出了一系列洞察，结合imperical studies的验证，对改进PIDSes提出了建议。例如，我们定性与定量的分析了当前PIDSes好坏的本质原因，我们分析了误报产生的原因并提出了缓解方法，我们提出了一些被忽视但重要的问题。
Based on the evaluation results and our industrial practices, we provide several insights and suggestions to solve the above problems and improve PIDSes. These primary insights are as follows:
\begin{itemize}[nosep]
% \begin{itemize}[nosep, leftmargin=*]
    \item We explain why PIDSes perform well or poorly from both qualitative and quantitative perspectives. In a word, we find that the primary factor affecting the performance is the feature learning task of a PIDS. The quantitative analysis shows that the Pearson correlation coefficient between the feature learning task and the AUC is approximately 0.73, with a p-value of 0.02.
    \item We analyze the causes of high false positive rates and propose a mitigation method. We category the causes into three types: sparse activities, unknown activities and semantic changes of activities. Our proposed method can reduce the false positive rate from about 25\% to about 10\%, and reduce the manual effort by 2/3.
    \item We find some problems may be overlooked but are important in practice. For example, because of the inconsistency between the target of fine-grained detection and that of investigation, high detection performance may also result in low investigation efficiency, exacerbating the alert fatigue problem.
\end{itemize} 

Finally, we provide several suggestions based on these insights. We hope that these insights and suggestions will help to improve existing PIDSes and guide future research.

In summary, our contributions are as follows:
\begin{itemize}[nosep]
% \begin{itemize}[nosep, leftmargin=*]
    \item We analyze the differences between the data from DARPA datasets and that collected in industrial scenarios, identifying three main new characteristics in industry.
    \item We build several datasets in industrial scenarios to evaluate the performance of five SOTA PIDSes. \textit{To the best of our knowledge, this is the first systematic evaluation of PIDSes in industrial scenarios.}
    \item We provide insights and suggestions based on the evaluation results and our industrial practices to solve the proposed problems and improve existing PIDSes.
\end{itemize}

The rest of the paper is organized as follows. We present the background and some neccessary knowledge in \S~\ref{sec:background}. Then, we introduce the research questions and the evaluation setup in \S~\ref{sec:setup}. The evaluation results are provided in \S~\ref{sec:measurement}. Based on the results, we provide insights and suggestions in \S~\ref{sec:insight}. \S~\ref{sec:discussion} discusses some other issues and the future work. We review related work on the survey of PIDSes in \S~\ref{sec:related-work} and conclude the paper in \S~\ref{sec:conclusion}.
\section{Background}
\label{sec:background}

\subsection{Introduction to PIDS}
\label{sec:intro-pids}

\begin{figure}[t]
    \centering
    \includegraphics[width=0.9\linewidth]{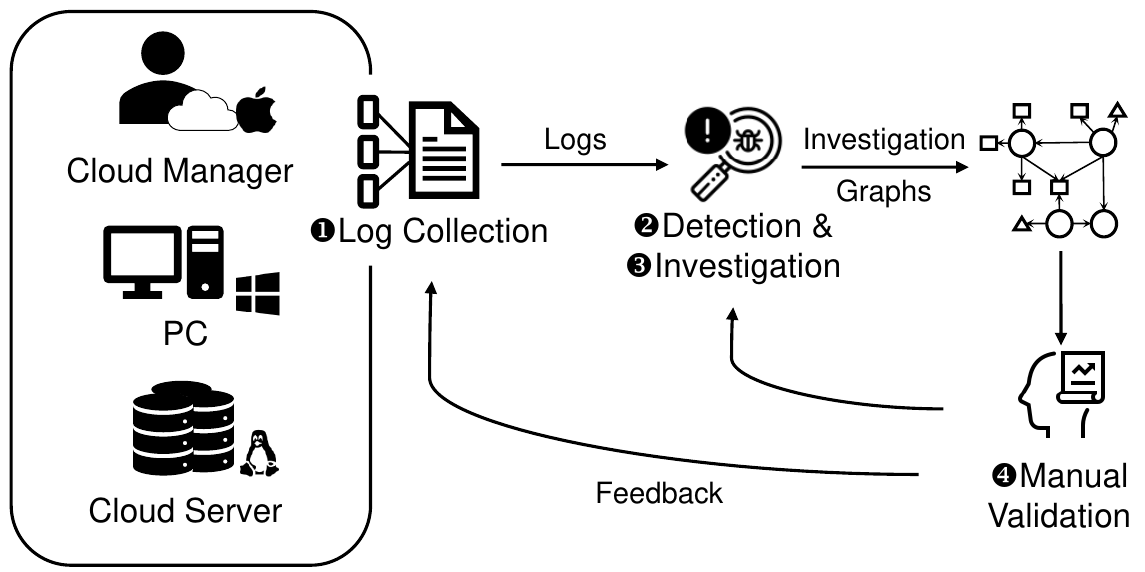}
    \caption{The pipeline of attack detection with a PIDS.}
    \label{fig:pipeline}
\end{figure}

PIDSes aim to detect attacks via system logs. In this subsection, we give a brief introduction to the PIDS.
Fig.~\ref{fig:pipeline} shows the whole pipeline of attack detection with a PIDS.
In general, a PIDS takes system logs as input and outputs a provenance graph composed of attack behaviors, then the analysts validate the results and give feedback.

\noindent \textbf{Log collection and protection.} In PIDSes, logs record the system activities, such as system calls. For example, when a process creates a file, there will be a log recording the creation event and the information of the process and the file.
A plenty of work focuses on collecting logs with low overhead during the collection and storage~\cite{Custos, linux-prov-capture, spade, hi-fi, ALASTOR, eaudit, palantir, LEONARD, nids-log-ebpf}.
Besides, to prevent attackers from tampering with logs, some work also focuses on log protection~\cite{logging-system, hardlog, KennyLoggings, logging-system}.

\noindent \textbf{Attack deteciton.} The main goal of PIDSes is to detect attacks. PIDSes typically build a provenance graph, extract features, conduct detection, and output alerts in the form of graphs, nodes or edges.
The detection methods contain two main categories~\cite{sok-pids}: rule-based approaches and anomaly-based approaches. Rule-based approaches~\cite{SLEUTH, SAQL, MORSE, HOLMES, Poirot, RapSheet, p-gaussian, CAPTAIN} usually apply expert-defined rules to generate alerts, such as assigning tags to nodes with specific attributes and using tag propagation to track the attack.
Anomaly-based approaches~\cite{deeplog, UNICORN, SIGL, StreamSpot, Log2vec, provtalk, PROVDETECTOR, SHADEWATCHER, THREATRACE, PROGRAPHER, NODLINK, DISTDET, KAIROS, R-CAID, FLASH, MAGIC, orthrus,slot, STGAN, llm-detection,meng2025knowhow} learn the features of benign activities and deem the activities that deviate from these as anomalies.
As the anomaly-based approaches can detect unknown attacks, they have become mainstream in recent years.

\noindent \textbf{Attack investigation.}
The output of attack detection is usually a subset of the whole attack, e.g., the most suspicious nodes or edges in the provenance graph. The analysts need to further investigate the alerts to figure out the complete attack.
The studies exploring attack detection may also provide simple investigation strategies.
What's more, some work concentrates on providing effective and human-friendly investigation of attacks~\cite{OmegaLog, NODOZE, ATLAS, MPI, WATSON, DEPIMPACT, Provg-searcher, ALchemist, uiscope, mci, PRIOTRACKER, DEEPCOM, TREC, Contexts}.

In this paper, we mainly focus on the attack detection of PIDSes, especially anomaly-based approaches, as they are popular and perform well.

\subsection{New Characteristics in Industry}
\label{sec:industrial-character}

\begin{figure}[t]
    \centering
    \includegraphics[width=0.7\linewidth]{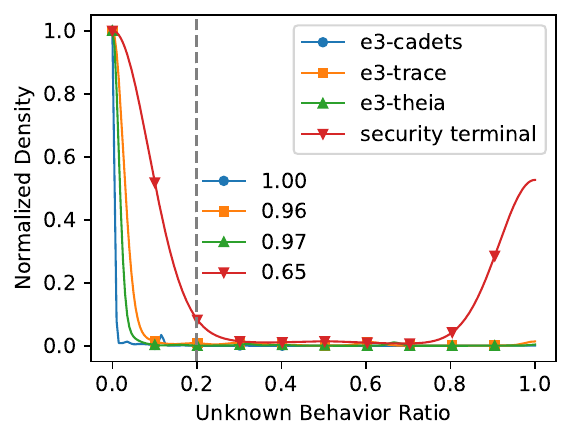}
    \caption{The distribution of unknown behaviors for each process in DARPA-E3 and a real-world host.}
    \label{fig:process-distr-compare}
\end{figure}

DARPA-TC datasets are widely used to evaluate PIDSes. However, we find that there are some new characteristics in industrial environment that are not well reflected in the DARPA-TC datasets. We summarize three main characteristics as follows.

\noindent \textbf{Heterogeneous multi-source inputs.} 
The number of hosts needed to be handled is very large, e.g., in our scenario, there are more than 5,000 security terminals and more than 10,000 cloud servers. 
If a PIDS is deployed on each host independently, it will lead to high storage and computation overhead. Maintaining PIDSes on each host would also be unaffordable.
What's more, the operating systems are heterogeneous, e.g., Windows and Linux. It is still unknown whether existing PIDSes handle such heterogeneous multi-source inputs well.

\noindent \textbf{More powerful attackers.} 
In practice, PIDSes may frace more powerful attackers. 
For example, the DARPA-TC datasets record all attack behaviors truthfully. However, we find attackers may invade the victim system before a PIDS is deployed. This means that the PIDS may incorrectly regard some attack behaviors as benign ones as they are already present in the previous logs. Furthermore, more techniques such as downloading through Tor network~\cite{tor-attack} are used to hide attack behaviors, making it more difficult for PIDSes to detect attacks.

% $\boldsymbol{\bullet}$ 
\noindent \textbf{Increasing benign activity complexity.}
In practice, user behaviour may change over time, which produces benign yet unknown activities. 
Anomaly-based PIDSes will deem such activities as anomalies, resulting in a high number of false positives.
We qualify the benign but unknown activities through a straightforward method. In a short, we calculate the proportion of unknown logs for each process, and the distribution of each process is shown in Fig.~\ref{fig:process-distr-compare}. We find that over 96\% of processes in the DARPA-E3 datasets have less than 20\% of unknown logs, whereas in reality, the ratio is about 65\% and about 20\% of processes have totally unknown logs. How these unknown logs affect the detection performance of PIDSes is still unknown.

\section{Measurement Setup}
\label{sec:setup}

\subsection{Research Questions}
Based on the new characteristics in industrial environment (as shown in \S~\ref{sec:industrial-character}), we aim to answer the following research questions via measurement:

% 需要被检测的主机十分多，理想的情况是使用某个主机的数据训练出的模型可以直接部署到其他主机上进行检测，从而节省大量的训练和维护成本。然而，这种跨主机和跨平台的可移植性在工业环境中尚未被充分研究。
\noindent \textbf{RQ1: What is the portability of PIDSes across multiple source inputs?} 
There are lots of hosts with different platforms and configurations that need to be protected. Ideally, we can train a PIDS on one host and deploy it on others, which can save training and maintenance costs. However, such portability across different hosts and different platforms has not been well studied. We will explore the portability of PIDSes in this research question.

\noindent \textbf{RQ2: What is the performance of PIDSes to detect real-world attacks?} 
In the real world, attackers are more capable. For example, some of attackers infiltrate the victim before any PIDS is deployed, and the trianing data may contain malicious behaviors, which violates the assumptions of the existing PIDSes. This research question will evaluate the performance of PIDSes when facing such powerful attacks.

\noindent \textbf{RQ3: What is the false positive rate of PIDSes, and what are the causes?} 
False positive is a long-standing issue for anomaly-based PIDSes. In real scenarios, some hosts have ever-changing activities (as shown in Fig.~\ref{fig:process-distr-compare}), which may lead to high false positives. This research question will measure the false positive rates of PIDSes and analyze their causes.

\noindent \textbf{RQ4: What is the time overhead of PIDSes?}
To deploy PIDSes in practice, the performance overhead should be acceptable, especially the time overhead. This research question will measure the time costs of PIDSes.

\subsection{Measurement Settings}

\noindent \textbf{Considered PIDSes.}
Anomaly-based PIDSes become the mainstream in recent years as their capability of detecting zero-day attacks. We consider five representative anomaly-based PIDSes in our evaluation, including node-level, process-level, and graph-level detection approaches.
(1) \magic\cite{MAGIC}: This approach applies graph attention networks to learn node embeddings and detect anomalous nodes via distances to benign nodes.
(2) \flash\cite{FLASH}: This approach converts anomaly detection to node type prediction on a graph, and uses a GraphSAGE~\cite{GraphSAGE} to learn node embeddings.
(3) \orthrus\cite{orthrus}: \orthrus  replies on edge prediction and a graph convolutional network to conduct node-level detection.
(4) \nodlink\cite{NODLINK}: Unlike the above three approaches, \nodlink learns the features of processes and detects anomalous processes via a constructed Steiner tree problem.
(5) \kairos\cite{KAIROS}: This approach uses a temporal graph network~\cite{TGN} and a multiple linear layer (MLP) to learn edge embeddings, get the anomaly score of each edge, and finally conduct graph-level detection.

\noindent \textbf{Implementation.}
We implement those PIDSes according to their open-source code, and we do not change their configurations, such as the hyperparameters and default threshold. We make sure that all the results claimed in their paper are reproducible under such configurations.
Although \kairos conducts graph-level deteciton, it provides the anomaly score for each edge. Thus, we can set a threshold to identify anomalous edges and further mark the corresponding nodes as malicious. After such modification, we can evaluate \kairos in the same way as the other node-level PIDSes.

\noindent \textbf{Data Collection.}
% 我们部署了Security Management与cloud workload protection服务，我们的数据来源于使用我们服务的用户。在Security Management服务中，用户系统为Windows系统，用户总数约为10^5台。在cloud workload protection服务中，用户系统为Linux系统，用户总数约为10^7台。在日志收集上，我们使用了ETW、Auditlog进行日志收集。
The data is collected from the users of two enterprise security services, the security management~\cite{security-management-standards} service and the cloud workload protection~\cite{cloud-workload-protection-survey} service. 
In the security management service, the user systems are Windows-based and the total number of users is about 5,000. In the cloud workload protection service, the user systems are Linux-based and the total number of users is about 10,000. 
We use ETW~\cite{ETW} and a modified version of Auditlog~\cite{LinuxAudit} for log collection.
All user information is anonymised, and we comply with relevant privacy policies and regulations.

\noindent \textbf{Dataset Construction and Groundtruth Labeling.}
% 我们的数据是海量的，我们需要从中选择一部分有代表性的数据构造数据集用于测试。我们开发了一套基于规则的系统进行攻击检测，将产生的告警进行人工分析，并选择有代表性的攻击及其所在日期的日志构成数据集。在获取数据集后，我们以实体的名称（如进程名、文件名等）作为判定标准，将对应的单个event与对应的node标记为恶意。
We need to select a part of representative data from the massive data to construct datasets for testing.
We develop a rule-based intrusion detection system to generate alerts, then manually analyze these alerts to investigate the entire attack procedure and select representative attacks. The logs on the dates when these attacks occur are used to construct datasets.
About the groundtruth labeling, we use the names of indicators of compromise (IOCs), such as process names and file names, to determine whether the corresponding edges and nodes are malicious.

\noindent \textbf{Metrics.}
We use true positive (TP), true negative (TN), false positive (FP), false negative (FN), true positive rate (TPR), and false positive rate (FPR) as the metrics. These metrics reflect the detection capability under a certain threshold.
Moreover, we record Area Under Curve (AUC) for each approach to evaluate the overall detection performance under different thresholds. A higher AUC indicates that the approach can distinguish more effectively between benign and malicious nodes, enabling analysts to detect attacks with fewer false alerts.
We record metrics in the node level for \magic, \flash, \orthrus, and \kairos, while in the process level for \nodlink.

\section{Measurement Results}
\label{sec:measurement}

In this section, we show the measurement results for the above research questions. We repeat the detection of each PIDS 10 times and report the average metrics.

\subsection{The portability of PIDSes across multiple source inputs (RQ1)}
\label{sec:measurement-portability}

\begin{figure}[t]
    \centering
    \subfloat[Security terminal (Windows).\label{fig:sanbox-windows}]{\includegraphics[width=0.23\textwidth]{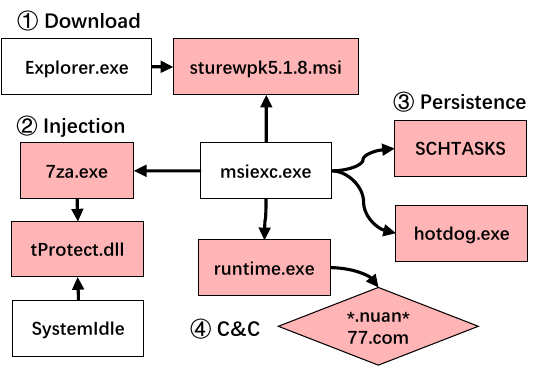}
    }
    \hfill 	
    \subfloat[Cloud host (Linux).\label{fig:sanbox-linux}]{\includegraphics[width=0.23\textwidth]{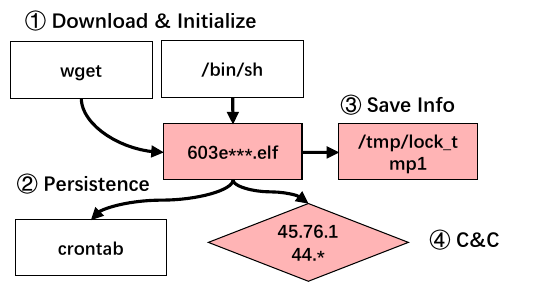}}
    \caption{The attack in the sandboxs on different platforms.}
    \label{fig:sandbox-attacks}
\end{figure}

% \begin{figure}[t]
%     \centering
%     \includegraphics[width=0.7\linewidth]{figures/sanbox_windows.pdf}
%     \caption{The attack in the security terminal (Windows) sandbox.}
%     \label{fig:sanbox-windows}
% \end{figure}

% \begin{figure}[t]
%     \centering
%     \includegraphics[width=0.7\linewidth]{figures/sandbox_linux.pdf}
%     \caption{The attack in the cloud host (Linux) sandbox.}
%     \label{fig:sanbox-linux}
% \end{figure}

\begin{table}[t]
\caption{Detection results on the security terminal (sandbox).}
\label{tab:eva-ideal-sec-host}
\resizebox{.99\linewidth}{!}{
\begin{tabular}{@{}cccccccc@{}}
\toprule
\textbf{Method} & \textbf{TPR} & \textbf{FPR} & \textbf{AUC}  & \textbf{TP} & \textbf{TN}    & \textbf{FN} & \textbf{FP} \\ \midrule
\magic & 0.7332 & 0.4510 & 0.7379 & 32.26 & 12940.94 & 11.74 & 10632.06 \\
\flash & 0.2591 & 0.0477 & 0.9004 & 11.40 & 22449.60 & 32.60 & 1123.40 \\
\orthrus & 0.0227 & 0.0001 & 0.9677 & 1.00 & 23570.12 & 43.00 & 2.88 \\
\kairos & 0.9545 & 0.1197 & 0.9437 & 42.00 & 20752.00 & 2.00 & 2821.00 \\
\nodlink & 0.8133 & 0.3576 & 0.7837 & 21.96 & 1897.62 & 5.04 & 1056.38 \\
\bottomrule
\end{tabular}
}
\end{table}

\begin{table}[t]
\caption{Detection results on the cloud host (sandbox).}
\label{tab:eva-ideal-cloud}
\resizebox{.99\linewidth}{!}{
\begin{tabular}{@{}cccccccc@{}}
\toprule
\textbf{Method} & \textbf{TPR} & \textbf{FPR} & \textbf{AUC}  & \textbf{TP} & \textbf{TN}    & \textbf{FN} & \textbf{FP} \\ 
\midrule
\magic & 0.3305 & 0.1883 & 0.5784 & 278.30 & 802.00 & 563.70 & 186.00 \\
\flash & 0.0019 & 0.0273 & 0.6254 & 1.60 & 961.00 & 840.40 & 27.00 \\
\orthrus & 0.0036 & 0.0101 & 0.1911 & 3.00 & 978.00 & 839.00 & 10.00 \\
\kairos & 0.2705 & 0.5065 & 0.2877 & 227.80 & 487.60 & 614.20 & 500.40 \\
\nodlink & 0.0181 & 0.0271 & 0.8464 & 15.20 & 921.30 & 823.80 & 25.70 \\
\bottomrule
\end{tabular}
}
\end{table}

We create an ideal environment for measurement, allowing us to concentrate on the influence of different source inputs. Specifically, we use a security terminal (Windows) sanbox and a cloud host (Linux) sandbox to collect training data and launch attacks. The training data only contains benign activities about system initialization.
What's more, we use a security terminal from a real user to collect training data for evaluating portability across different hosts.

The details of attacks are shown in Fig.\ref{fig:sanbox-windows} and Fig.\ref{fig:sanbox-linux}. These two attacks both aim to inject a backdoor on the victim's system via malware. Then, attackers can steal sensitive information from the victim or conduct further attacks.

\noindent \textbf{Detection results in the ideal environment.}
The average metrics are shown in Table~\ref{tab:eva-ideal-sec-host} and Table~\ref{tab:eva-ideal-cloud}.
From the results, we can see that the detection results differ among different PIDSes. In general, \orthrus and \flash adopt more conservative detection strategies, thereby have low TPRs and FPRs. Meanwhile \magic, \kairos, and \nodlink have higher TPRs with higher FPRs.
Compare the two platforms, all the PIDSes perform better on the security terminal than on the cloud host (i.e., get higher AUCs). 
One possible reason is that, in the cloud host, the data is imbalanced with a majority of process nodes and the edges of process opening and closing. The imbalanced data reduces the difference between benign and malicious activities, making detection more difficult. Also, the representation learning of each method is ineffective when using such training data.
On the other hand, the attacker on cloud host only installs the backdoor and connects to the C2 server a few times, making its behaviour stealthy.

\begin{table}[t]
\caption{Detection results across different hosts.}
\label{tab:eva-cross-host}
\resizebox{.99\linewidth}{!}{
\begin{tabular}{@{}cccccccc@{}}
\toprule
\textbf{Method} & \textbf{TPR} & \textbf{FPR} & \textbf{AUC}  & \textbf{TP} & \textbf{TN}    & \textbf{FN} & \textbf{FP} \\ 
\midrule
\magic & 0.4741 & 0.5531 & 0.5366 & 20.86 & 10535.68 & 23.14 & 13037.32 \\
\flash & 0.3645 & 0.8308 & 0.2632 & 16.04 & 3989.30 & 27.96 & 19583.70 \\
\orthrus & 0.0441 & 0.0005 & 0.8351 & 1.94 & 23561.54 & 42.06 & 11.46 \\
\kairos & 0.3750 & 0.0407 & 0.5520 & 16.50 & 23573.00 & 27.50 & 1000.50 \\
\nodlink & 0.6481 & 0.1902 & 0.8081 & 17.50 & 2392.02 & 9.50 & 561.98 \\
\bottomrule
\end{tabular}
}
\end{table}

% AUC decrease: 0.2013, 0.6372, 0.1326, 0.3917, -0.0244, average: 0.2677, from -0.0244 to 0.6372

\begin{table}[t]
\caption{Detection results from Linux to Windwos.}
\label{tab:eva-cross-platform-l2w}
\resizebox{.99\linewidth}{!}{
\begin{tabular}{@{}cccccccc@{}}
\toprule
\textbf{Method} & \textbf{TPR} & \textbf{FPR} & \textbf{AUC}  & \textbf{TP} & \textbf{TN}    & \textbf{FN} & \textbf{FP} \\ 
\midrule
% magic & 0.9236 & 0.9932 & 0.6297 & 40.64 & 161.28 & 3.36 & 23411.72 \\
% flash & 0.3668 & 0.7434 & 0.2788 & 16.14 & 6047.86 & 27.86 & 17525.14 \\
% orthrus & 0.0682 & 0.0000 & 0.5098 & 3.00 & 23489.92 & 41.00 & 1.08 \\
% nodlink & 0.4370 & 0.4954 & 0.4549 & 11.80 & 1490.58 & 15.20 & 1463.42 \\
magic & 0.9302 & 0.9938 & 0.6645 & 40.93 & 147.20 & 3.07 & 23425.80 \\
flash & 0.3409 & 0.5436 & 0.3116 & 15.00 & 10758.00 & 29.00 & 12815.00 \\
orthrus & 0.0000 & 0.0002 & 0.7219 & 0.00 & 23487.00 & 44.00 & 4.00 \\
kairos & 0.6591 & 0.4705 & 0.2310 & 29.00 & 23491.00 & 15.00 & 20877.00 \\
nodlink & 0.0444 & 0.0309 & 0.5027 & 1.20 & 2862.70 & 25.80 & 91.30 \\
\bottomrule
\end{tabular}
}
\end{table}

% AUC decrease: 0.7379 - 0.6645 = 0.0734, 0.9004 - 0.3116 = 0.5888, 0.9677 - 0.7219 = 0.2458, 0.9437 - 0.2310 = 0.7127, 0.7837 - 0.5027 = 0.2810, average: 0.3803, from 0.0734 to 0.7127

\begin{table}[t]
\caption{Detection results from Windows to Linux.}
\label{tab:eva-cross-platform-w2l}
\resizebox{.99\linewidth}{!}{
\begin{tabular}{@{}cccccccc@{}}
\toprule
\textbf{Method} & \textbf{TPR} & \textbf{FPR} & \textbf{AUC}  & \textbf{TP} & \textbf{TN}    & \textbf{FN} & \textbf{FP} \\ 
\midrule
magic & 0.9907 & 0.7644 & 0.7436 & 834.20 & 232.80 & 7.80 & 755.20 \\
flash & 0.0026 & 0.0308 & 0.9335 & 2.20 & 957.60 & 839.80 & 30.40 \\
orthrus & 0.0012 & 0.0051 & 0.4065 & 1.00 & 983.00 & 841.00 & 5.00 \\
kairos & 1.0000 & 0.4936 & 0.6009 & 842.00 & 988.00 & 0.00 & 963.00 \\
nodlink & 0.9996 & 0.8378 & 0.5235 & 838.70 & 153.60 & 0.30 & 793.40 \\
\bottomrule
\end{tabular}
}
\end{table}

\noindent \textbf{Detection results across different hosts.}
We use the data from a real security terminal user for training and test it with the data in the security terminal sandbox. We confirm that there is no attacks in the training data. The results are shown in Table~\ref{tab:eva-cross-host}. 
When trained with data from another host, the TPRs and AUCs of most PIDSes descrease compared to Table~\ref{tab:eva-ideal-sec-host}. 
The reason is that there are more activities in training data, such as network connection and file downloading. The attacker also carries out these activities, which makes the difference between benign and malicious activities smaller.

\noindent \textbf{Detection results across different platforms.}
We exchange the training data of the two sandboxs to evaluate the portability across different platforms.
The results are shown in Table~\ref{tab:eva-cross-platform-l2w} and Table~\ref{tab:eva-cross-platform-w2l}.
When training on the cloud host and testing on the security terminal (Table~\ref{tab:eva-cross-platform-l2w}), all PIDSes have lower AUCs, and their TPRs are very close to FPRs. The reason is that the data on the two platforms differ significantly. Compared to the activities on one platform, the activities on the other platform are all unfamiliar and difficult to distinguish. 
Similar to the above results, when training on the security terminal and testing on the cloud host (Table~\ref{tab:eva-cross-platform-w2l}), all PIDSes have similar TPRs and FPRs. 
However, the AUCs of \magic, \flash, and \kairos increase. The reason is that the data on the security terminal is more balanced and diverse, which helps the PIDSes to learn more comprehensive features of benign activities.

\noindent \textbf{Summary.}
The portability of PIDSes across different source inputs is limited. When the training and testing data are from different hosts or platforms, the detection performance degrades significantly.

\subsection{The performance of PIDSes to detect real-world attacks (RQ2)}

We select two real-world attacks found by us for evaluation. The first one is a malicious mining attack on the cloud, which is a typical attack on cloud and uses various stealth techniques. The second one is a special long-term information stealing attack, where the attacker has already infiltrated the victim before any PIDS is deployed.

\begin{figure}[t]
    \centering
    \includegraphics[width=0.85\linewidth]{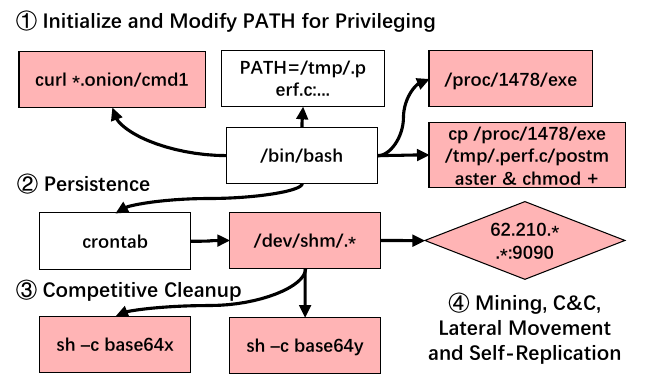}
    \caption{An attack utilises known vulnerabilities in postgreSQL to launch mining.}
    \label{fig:real-case-mining-attack}
\end{figure}

\begin{table}[t]
\caption{Detection results for the mining attack.}
\label{tab:eva-real-case-mining}
\resizebox{.99\linewidth}{!}{
\begin{tabular}{@{}cccccccc@{}}
\toprule
\textbf{Method} & \textbf{TPR} & \textbf{FPR} & \textbf{AUC}  & \textbf{TP} & \textbf{TN}    & \textbf{FN} & \textbf{FP} \\ 
\midrule
\magic & 0.3949 & 0.5142 & 0.3943 & 12528.63 & 109931.10 & 19200.37 & 116373.90 \\
\flash & 0.0035 & 0.0427 & 0.5375 & 111.00 & 216637.40 & 31618.00 & 9667.60 \\
\orthrus & 0.0003 & 0.0000 & 0.6419 & 9.00 & 226296.00 & 31720.00 & 9.00 \\
\kairos & 0.0163 & 0.5237 & 0.5577 & 3427.00 & 25744.00 & 206981.00 & 28302.00 \\
\nodlink & 0.8617 & 0.1850 & 0.9217 & 27243.60 & 175477.80 & 4372.40 & 39830.20 \\
\bottomrule
\end{tabular}
}
\end{table}

% AUCs: 0.3943, 0.5375, 0.6419, 0.5577, 0.9217, average: 0.6106, from 0.3943 to 0.9217

\noindent \textbf{Malicious Mining on the Cloud.}
We discover a large-scale attack that aims to compromise cloud hosts to conduct mining. The attack affected over 2,000 cloud hosts. This attack is typical of those targeting the cloud, involving intrusion, execution, and persistence. What's more, the attack uses lots of strategies to keep stealthy.
Thus, we choose it as a case for evaluation.

The details of the attack is shown in Fig.\ref{fig:real-case-mining-attack}.
First, it utilises weak password to get access of PostgreSQL and uses the PROGRAM feature to execute commands.
Then, it downloads malware via the Tor network and modifies the PATH variable to disguise the malware as a legitimate postmaster command.
For persistence, the attacker modifies the crontab to schedule its periodic execution.
Finally, the attack uses in-memory execution to clean up other potential mining programs, connect to a mining pool, and conduct self-replication.
The executed command lines are always encoded with Base64 to hide their semantic information. The in-memory execution can utilize existing processes to execute and suppress telemetry~\cite{in-memory}, making detection more difficult.

We randomly select a cloud host compromised by this attack and collect its logs from the first day of the attack for testing. We use the 4-day logs prior to the attack for training. The detection results are presented in Table~\ref{tab:eva-real-case-mining}.
From the results, we can see that most PIDSes have low AUCs (lower than 0.65) and only detect a small proportion of malicious nodes, except \nodlink.
The reason is similar to the sandbox results: the data is highly imbalanced with a large number of process nodes. Also, strategies such as Base64 encoding make attacks difficult to detect.
On the contrary, the malicious processes involving Base64 command lines are so uncommon that \nodlink can effectively identify them (86.17\% TPR and 92.17\% AUC).

\begin{figure}[t]
    \centering
    \includegraphics[width=0.85\linewidth]{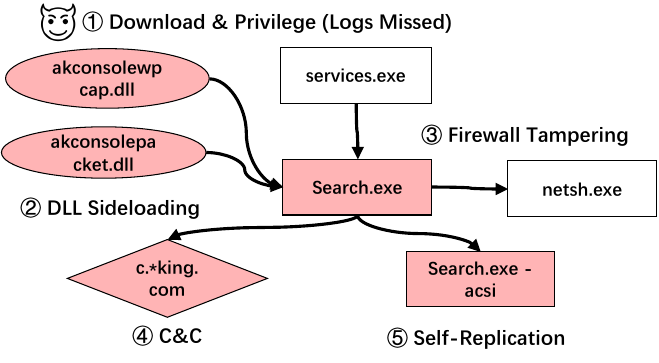}
    \caption{An attack invades the victim before any PIDS is deployed and tries to steal the sensitive information.}
    \label{fig:real-case-long-term-attack}
\end{figure}

\begin{table}[t]
\caption{Detection results for the information stealing attack.}
\label{tab:eva-real-case-stolen}
\resizebox{.99\linewidth}{!}{
\begin{tabular}{@{}cccccccc@{}}
\toprule
\textbf{Method} & \textbf{TPR} & \textbf{FPR} & \textbf{AUC}  & \textbf{TP} & \textbf{TN}    & \textbf{FN} & \textbf{FP} \\ 
\midrule
\magic & 0.1919 & 0.3475 & 0.4216 & 3.07 & 13518.50 & 12.93 & 7199.50 \\
\flash & 0.2500 & 0.2955 & 0.4855 & 4.00 & 14596.40 & 12.00 & 6121.60 \\
\orthrus & 0.0000 & 0.0008 & 0.5399 & 0.00 & 20702.00 & 16.00 & 16.00 \\
\kairos & 0.1250 & 0.2568 & 0.4840 & 2.00 & 15398.00 & 14.00 & 5320.00 \\
\nodlink & 0.9500 & 0.1270 & 0.9372 & 11.40 & 10613.60 & 0.60 & 1543.40 \\
\midrule
\multicolumn{8}{c}{{Training with data in the security terminal sandbox}} \\
\midrule
\magic & 0.7794 & 0.7797 & 0.4452 & 12.47 & 4565.10 & 3.53 & 16152.90 \\
\flash & 0.2625 & 0.3915 & 0.4252 & 4.20 & 12606.20 & 11.80 & 8111.80 \\
\orthrus & 0.0000 & 0.0003 & 0.6704 & 0.00 & 20711.50 & 16.00 & 6.50 \\
\kairos & 1.0000 & 0.9475 & 0.5714 & 16.00 & 1087.00 & 0.00 & 19631.00 \\
\nodlink & 0.7333 & 0.7469 & 0.5544 & 8.80 & 3076.87 & 3.20 & 9080.13 \\
\bottomrule
\end{tabular}
}
\end{table}

% AUCs: 0.4216, 0.4855, 0.5399, 0.4840, 0.9372, average: 0.5736, from 0.4216 to 0.9372

\noindent \textbf{Long-Term Information Stealing Attack.}
It is important for anomaly-based PIDSes to obtain benign activities as training data and references for detection. However, this requirement may not be met in industry. 
We encounter a long-term attacker who has already infiltrated the victim's host before any PIDS is deployed. Thus, the whole procedure about how the attacker invades the system and downloads malware is unknown. The training data in this case contains both benign and malicious activities.

The details of the attack is shown in Fig.\ref{fig:real-case-long-term-attack}.
First, the attacker downloads multiple tools and gets privileges in an unknown way.
In execution, the attack uses program "C:\verb|\\|Windows\verb|\\|System\verb|\\|\verb|@|xxx\verb|\\|Search.exe" to load corresponding .dll files and establishes network connection with the malicious websites to send the sensitive information. At last, the attack replicates itself for persistent concealment.
The activities involved in this attack resemble some benign network activities, i.e., it is common for a process to load .dll files and establish network connections.

We use 4-day logs for training and 1-day logs for testing. The training data contains both benign and malicious activities. We also use the logs in the security terminal sandbox as training data for comparison. The results are shown in Table~\ref{tab:eva-real-case-stolen}.
The results are similar to those in Table~\ref{tab:eva-real-case-mining}. The stealthy attack is difficult to detect. Most of PIDSes cannot distinguish attacks from benign activities.
In \nodlink, although the attack behaviors are present in the training data, \nodlink still generates alarms. This is because the attack behaviors are still in the minority compared to benign activities and are therefore treated as anomalies.
As for training with data from the sandbox, significant differences between the training and testing data produce higher TPs and FPs.

\noindent \textbf{Summary.}
The performance of PIDSes to detect real-world attacks is inadequate. This is due to both the training data and the stealth of the attacks themselves.

\subsection{The false positive rate of PIDSes on a host with ever-changing activities (RQ3)}
\label{sec:measurement-fpr}

\begin{figure}[t]
    \centering
    \includegraphics[width=0.65\linewidth]{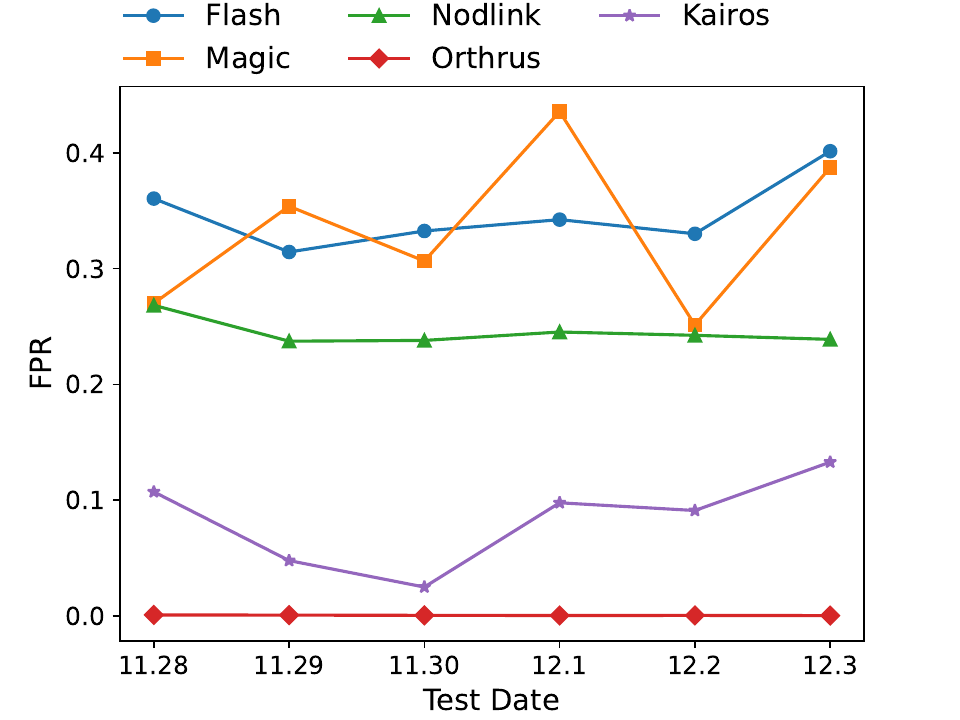}
    \caption{The FPRs of PIDSes on a security terminal with ever-changing activities.}
    \label{fig:real-case-ever-changing}
\end{figure}

False positive of PIDSes is a long-standing issue. In practice, we find that some hosts have ever-changing activities every day, which may lead to high FPRs. 
We analyze one such host, which unknown processes appear every day (e.g., unknown file names, command line parameters, and IP addresses).
We use the data on 11/27 as the training set and test it with the data for the next week. The results are shown in Fig.\ref{fig:real-case-ever-changing}.

The FPRs of \orthrus are pretty low (less than 1\%) because it adopts a conservative detection strategy to reduce the number of alerts.
The FPRs of \kairos are also low (less that 12\%) as the changes to the edges are insignificant in this host.
The FPRs of the other PIDSes are higher than 23\%. This means that these PIDSes may generate FPs frequently without any attacks, overwhelming analysts with false alerts.

We analyze these false positives and find that they can be divided into three categories: 
(1) Sparse activities: these activities have already appeared in the training data, but they are inherently rare, which leads to false positives.
This is the majority (about 50\%) of the false positives in \magic and \flash.
Specifically, the process svchost.exe hosts over 2,000 scripts for collecting hardware information, which is uncommon. As a result, all the script nodes become false positives.
(2) Unknown activities: these activities are not present in the training data and therefore PIDSes cannot learn their features. For example, the user download a large number of images during an online meeting, and there is a process node corresponded to several network nodes and file nodes. As this behavior has never occurred before, it results in false positives.
(3) Semantic changes: these activities means the changes of semantic information, such as file names, command line parameters, and IP addresses.
The PIDSes that use semantic information for embedding may be affected, e.g., the false positives in \nodlink are primarily caused by this.

\noindent \textbf{Summary.}
PIDSes will generate many false positives, especially on hosts with ever-changing activities. We categorise the causes of false positives into three types: sparse activities, unknown activities, and semantic changes. We will explore how to reduce false positives in \S~\ref{sec:insight-fp-reduction}.

\subsection{Time overhead of PIDSes (RQ4)} 

\begin{figure}[t]
    \centering
    \subfloat[Training time of PIDSes.\label{fig:train-time}]{\includegraphics[width=0.23\textwidth]{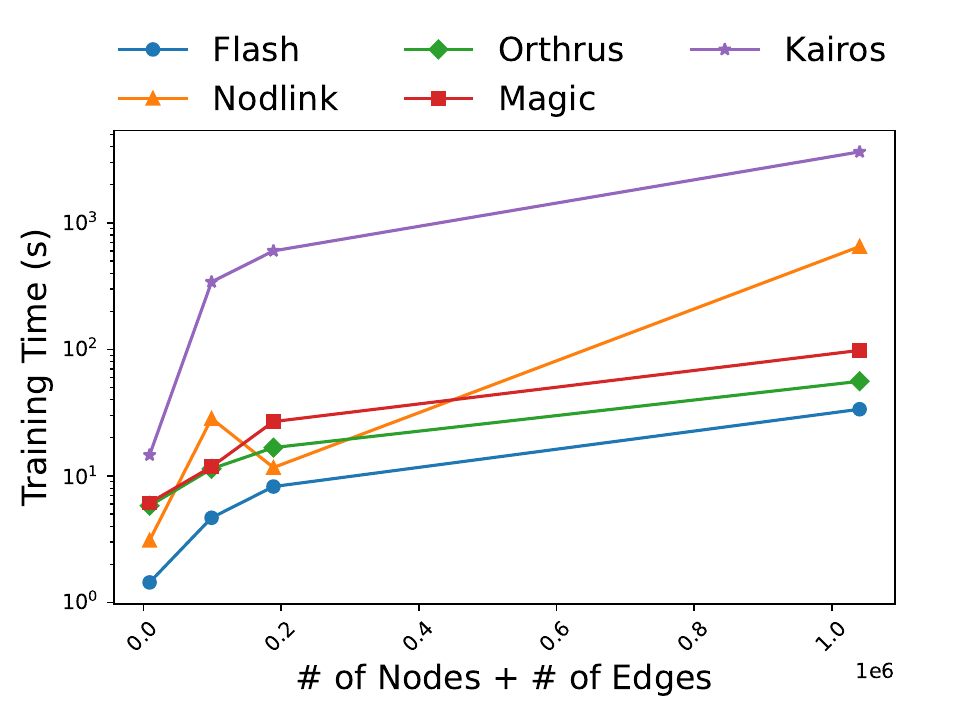}
    }
    \hfill 	
    \subfloat[Testing time of PIDSes.\label{fig:test-time}]{\includegraphics[width=0.23\textwidth]{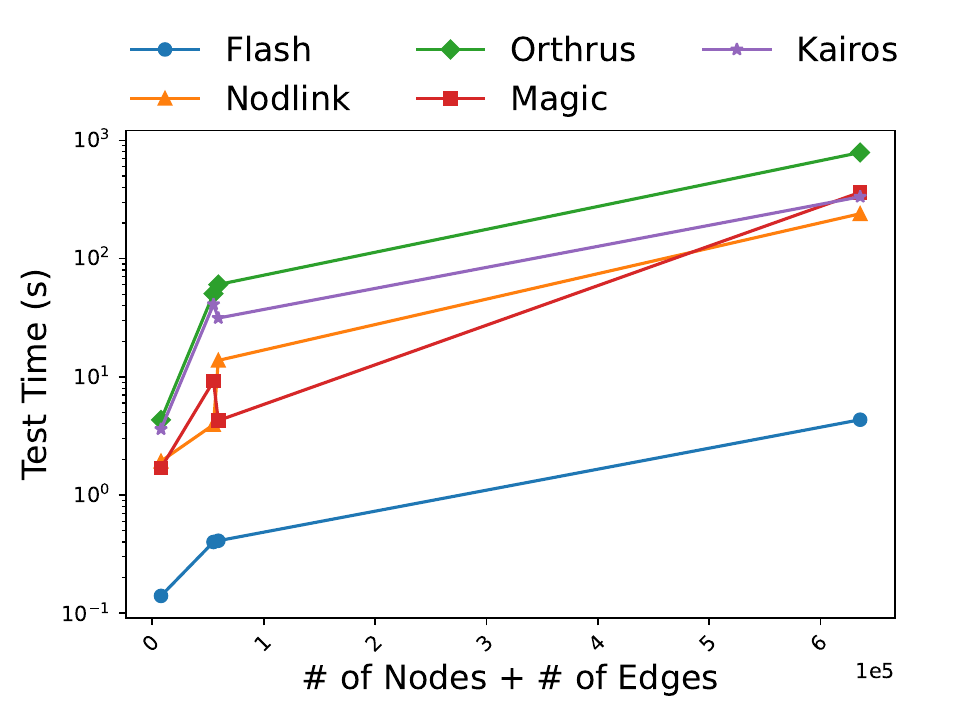}}
    \caption{The time costs under different graph sizes.}
    \label{fig:overhead}
\end{figure}

To deploy PIDSes in practice, the performance overhead should be acceptable, especially the time overhead. We measure the time costs of each PIDS under different graph sizes (the number of nodes and edges). The PIDSes are trained and tested on a 40GB A100 GPU. We exclude the time for data preprocessing and graph construction, focusing on the training and testing time.
The training time and testing time are shown in Fig.\ref{fig:train-time} and Fig.\ref{fig:test-time}.

% 结论是：不同方案的时间均随着图的规模线性增长。这符合大多数GNN的时间复杂度分析结果。各个方案的时间差异较大，甚至有数量级的差异。这是因为不同PIDSes的模型结构、表征维度、优化策略等均不同。我们注意到\flash的时间开销最低，在图规模为10^6时，其训练时间为33.65s，推理时间在规模超过6*10^5时为4.33s。这表明通过改变模型结构和优化策略，可以显著降低时间开销。
The results show that the time costs of all PIDSes increase linearly with the graph size, which is consistent with the time complexity analysis of most GNNs.
For different PIDSes, there are significant differences in time costs, sometimes by orders of magnitude. This is because different PIDSes have different model structures, representation dimensions, optimization strategies, and so on.
We note that \flash has the lowest time overhead. When the graph size is about $10^6$, its training time is 33.65s. Its testing time is 4.33s when the graph size exceeds $6\times10^5$. This indicates that changing the PIDS design could significantly reduce the time overhead.

\noindent \textbf{Summary.}
The time overhead of PIDSes increases linearly with the graph size. 
Different PIDSes have significantly different time costs. Although some PIDSes have high time overhead, we believe that it can be reduced by changing the model structure, configurations and optimization strategies.

\section{Insights, Suggestions, and Future Directions}
\label{sec:insight}

In this section, we summarize several insights obtained from our measurements, provide corresponding suggestions, and discuss future research directions for PIDSes.

\subsection{How can we improve the deteciton performance of PIDSes?}

% \begin{figure}[t]
%     \centering
%     \includegraphics[width=0.65\linewidth]{figures/gnn_analyzer.pdf}
%     \caption{The correlation between FPR/TPR and the distance of benign/malicious nodes among different datasets. As the distance increases, the benign activities are more likely to anomalies, leading to a higher FPR/TPR ratio.}
%     \label{fig:insight1-gnn-analyzer}
% \end{figure}

\begin{figure}[t]
    \centering
    \subfloat[Distance ratio with TPR/FPR.\label{fig:distance_tpr-fpr}]{\includegraphics[width=0.23\textwidth]{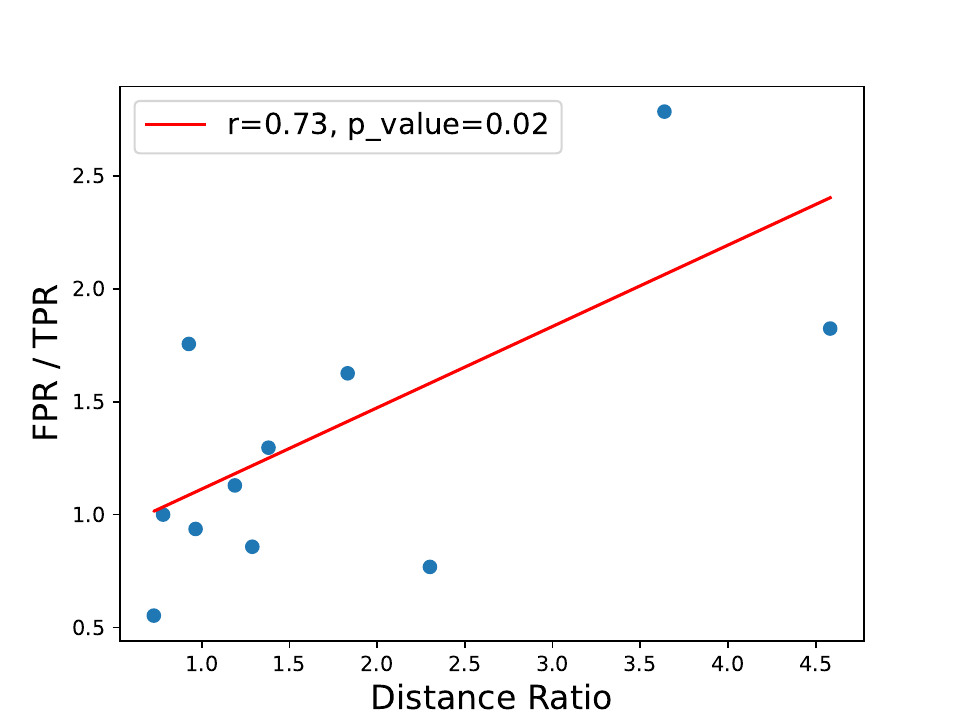}
    }
    \hfill 	
    \subfloat[Distance ratio with AUC.\label{fig:distance_auc}]{\includegraphics[width=0.23\textwidth]{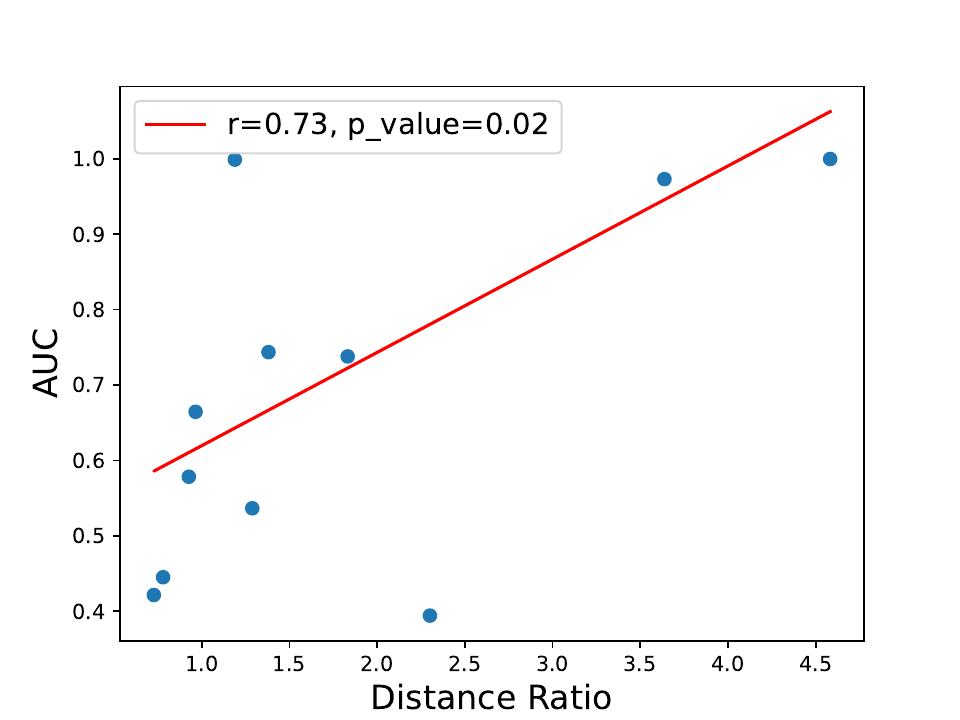}}
    \caption{The correlation between the distance ratio of malicious/benign nodes and and AUC. As the distance ratio increases, the malicious activities are more far away from the benign ones, leading to a higher TPR/FPR and a higher AUC.}
    \label{fig:gnn_analyzer}
\end{figure}

\noindent \textbf{Insight 1: Feature learning tasks are the primary factor influencing detection results.}
In our evaluation, we observe that the detection results of PIDSes vary under different settings. 
Many factors can influence the detection results, including the used data, model training, threshold setting, and so on.
Among these factors, we find that the primary factor influencing the detection results of PIDSes is the feature learning tasks themselves.

The feature learning task refers to the task used by a PIDS to obtain embeddings of nodes (or other types of entities). 
If these tasks can reflect the differences between malicious and benign nodes, PIDSes will achieve great detection results; otherwise, it will be difficult to balance false positives and false negatives.
For example, \nodlink achieves high performance in our datasets because the malicious processes differ greatly from the benign ones, whereas the benign processes have stable semantic information, resulting in high TPRs with low FPRs.

In order to further validate this insight, we also conduct a quantitative analysis. We take \magic as an example for analysis.
\magic uses GAT for node embedding learning, which is a representative type of GNN. 
Moreover, \magic does not use semantic information for node representation,  meaning we can eliminate the uncertainty caused by semantic representation.

\magic designs a node reconstruction task and an edge prediction task for feature learning.
The two tasks are related to the features of neighboring nodes and edge types.
Therefore, if there are differences between malicious and benign nodes in terms of the features of neighboring nodes and edge types, \magic may achieve great detection results.
We count the type information of neighboring nodes and edges for each node, and represent a node as a vector.
Then, we calculate the distance between nodes in the testing set and nodes in the training set based on these vectors. Details are provided in the Appendix~\ref{sec:appendix-distance}.
A larger distance indicates that malicious nodes are more different from benign nodes in terms of the features of neighboring nodes and edge types. \magic will perform better with a larger distance.

We evaluate the correlation between the ratio of distances of malicious/benign nodes and the ratio of TPR and FPR, also the correlation between the distance ratio and the AUC. We calculate the results on different datasets, including the DARPA-E3 datasets, the datasets usd in Table~\ref{tab:eva-ideal-sec-host}, Table~\ref{tab:eva-ideal-cloud}, Table~\ref{tab:eva-cross-host}, Table~\ref{tab:eva-cross-platform-l2w}, Table~\ref{tab:eva-cross-platform-w2l}, and Table~\ref{tab:eva-real-case-mining}.
The results are shown in Fig.~\ref{fig:gnn_analyzer}.
As can be seen, there is a strong correlation between the distance ratio and the TPR/FPR (when we remove one outlier point, the Pearson correlation coefficient is 0.71 with a 0.02 p-value). The result for AUC is similar.
The resluts indicate that when the differences between malicious and benign nodes in terms of the features of neighboring nodes and edge types are more significant, \magic can achieve a higher TPR with a lower FPR, leading to a higher AUC.

\noindent \textbf{Suggestion:} To achieve better detection results, we recommend to discover the differences between malicious and benign activities first, and then design feature learning tasks that can reflect these differences.
In this way, it can help PIDSes to achieve a high TPR with a low FPR.

% dataset & TP & TN & FN & FP & M2T & B2T \\
% e3-trace & 67382.60 & 278281.40 & 0.40 & 337710.60 & 51.52 & 11.24 \\
% e3-cadets & 12852.00 & 134729.40 & 0.00 & 75471.60 & 31.22 & 8.58 \\
% e3-theia & 22825.40 & 33985.40 & 0.20 & 263613.20 & 138.71 & 116.73 \\
% Table~\label{tab:eva-ideal-sec-host} & 32.26 & 12940.94 & 11.74 & 10632.06 & 36.90 & 20.15 \\
% Table~\label{tab:eva-cross-platform-l2w} & 40.64 & 161.28 & 3.36 & 23411.72 & 51.73 & 53.65 \\
% Table~\label{tab:eva-ideal-cloud} & 279.02 & 1291.76 & 33.98 & 225.24 & 108.79 & 86.54 \\
% Table~\label{tab:eva-cross-machine} & 20.86 & 10535.68 & 23.14 & 13037.32 & 38.28 & 29.73 \\
% Table~\label{tab:eva-cross-platform-w2l} & 290.70 & 38.80 & 1.30 & 1499.20 & 97.24 & 79.85 \\
% Table~\label{tab:eva-ideal-sec-host} & 3.07 & 13518.50 & 12.93 & 7199.50 & 14.67 & 20.20 \\
% Table~\label{tab:eva-ideal-sec-host}(below) & 12.47 & 4565.10 & 3.53 & 16152.90 & 24.03 & 30.84 \\
% Table~\label{tab:eva-real-case-mining} & 12528.63 & 109931.10 & 19200.37 & 116373.90 & 21.33 & 9.27 \\

\begin{figure}
    \centering
    \includegraphics[width=0.95\linewidth]{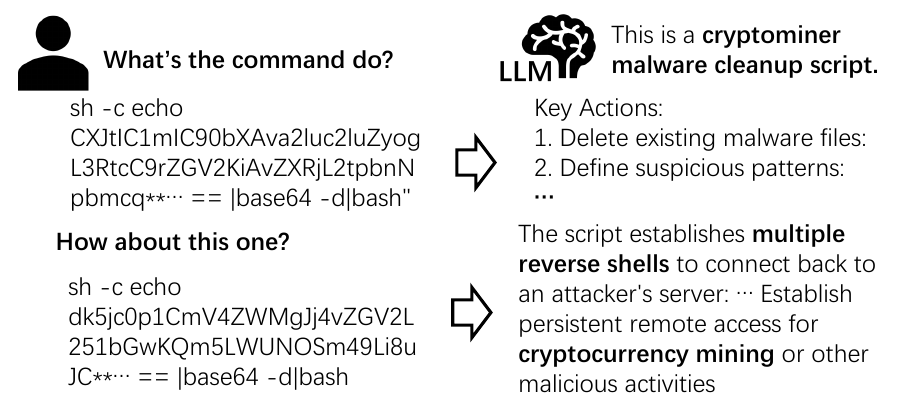}
    \caption{An example of using LLM to analyze the command lines of the mining attack.}
    \label{fig:insight1-semantic_usage}
\end{figure}

\noindent \textbf{Insight 2: Semantic information is important for attack detection.}
In our evaluation, we find that PIDSes that consider semantic information often achieve better detection results.
For example, \nodlink concentrates on the semantic differences between malicious and benign processes, achieving high AUC scores in most datasets.
In the GNN-based PIDSes, \flash, \orthrus, and \kairos use command lines, file names, and ip addresses for node embeddings, achieving higher AUC scores compared to \magic.

The evaluation results indicate that semantic information is important for attack detection, i.e., attacks often have different content compared with benign activities.
More importantly, some semantic information can be used directly to detect attacks, including process command lines, file names and website domain names. 
This is helpful for security analysts, as they can quickly identify an attack based on this semantic information (e.g., command lines) and then investigate the attack further.

The question of whether this semantic information can be used to automatically detect attacks in the same way that humans do remains open.
Fortunately, we find that large language models (LLMs) have the potential to help with this task.
Fig.~\ref{fig:insight1-semantic_usage} shows an example of using an LLM to analyze the command lines of the mining attack shown in Fig.~\ref{fig:real-case-mining-attack}.
The LLM first decodes the Base64-encoded shell into a human-readable text. Then, it analyzes the functionality of the shell step by step, pointing out that the command downloads a known mining script. At the same time, the LLM extracts key information such as the script name and C\&C server address. Finally, the LLM provides security recommendations.
Based on the capabilities of LLMs, we design an LLM agent for automatic attack analysis. When providing more external information, such as threat intelligence and different entity information, the LLM can give a clearer analysis of the attack.

\noindent \textbf{Suggestion:} We recommend to utilize semantic information for attack detection, and more fine-grained approaches are needed instead of simply getting word/sentence embeddings.
What's more, detecting attacks based on semantic information and external knowledge is a promising way to achieve training-free detection.

\subsection{How can we reduce false positives?}
\label{sec:insight-fp-reduction}

\begin{figure}[t]
    \centering
    \subfloat[FPR before/after reduction.\label{fig:insight-fpr}]{\includegraphics[width=0.23\textwidth]{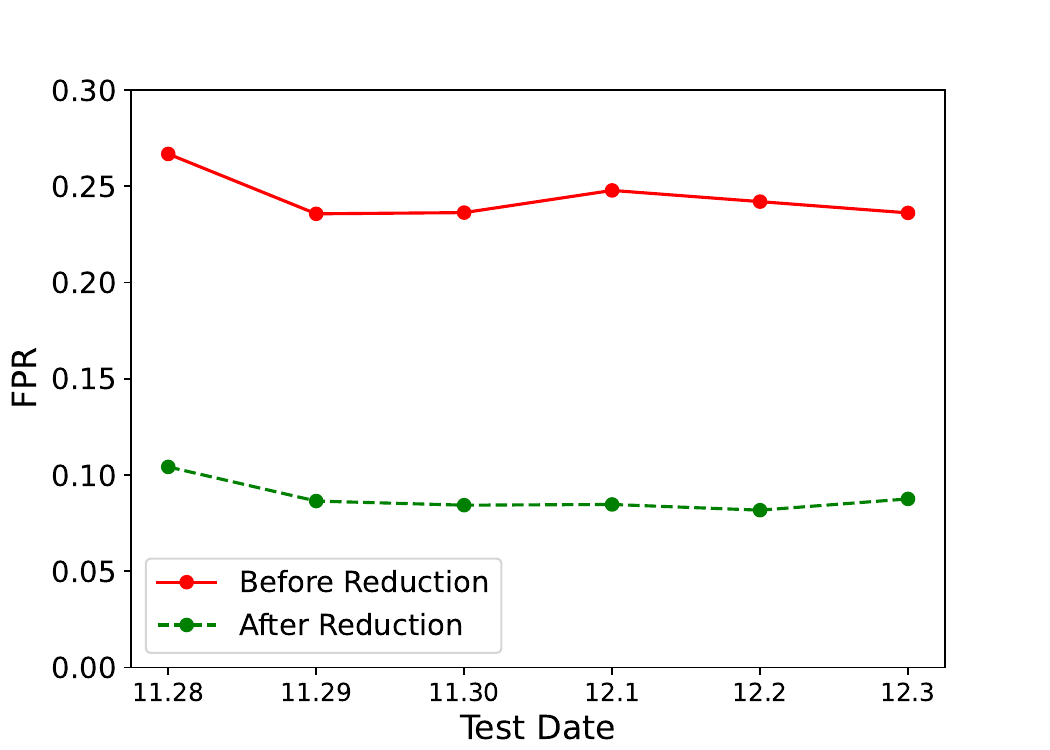}
    }
    \hfill 	
    \subfloat[\# of FPs and communities.\label{fig:insight-fp_num}]{\includegraphics[width=0.23\textwidth]{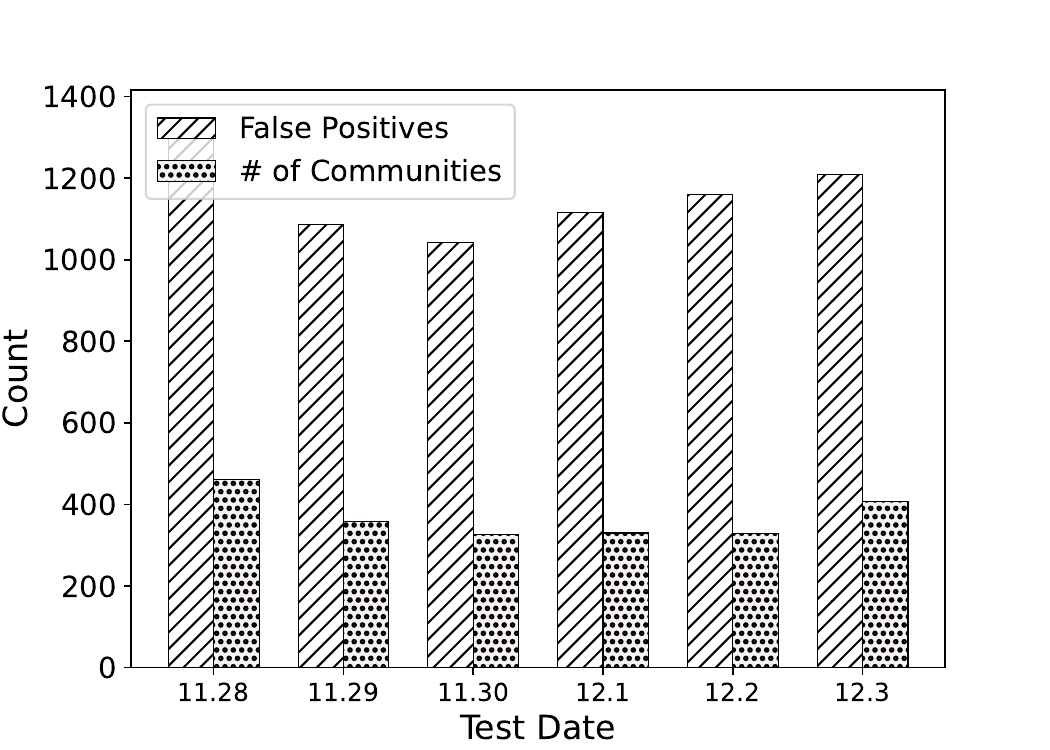}}
    \caption{Results of FP reduction for \nodlink via an unsupervised approach.}
    \label{fig:insight-fp_reduction}
\end{figure}

\noindent \textbf{Insight 3: Using unsupervised approaches may reduce false positive rates.}
PIDSes often raise a large number of false positives during detection.
In some cases, although the false positive rate seems not high, the number of false positives is still large, which poses a challenge for security analysts.
For example, in Table~\ref{tab:eva-real-case-mining}, \nodlink has a false positive rate of 18.5\%, resulting in 39,830 false positives that need to be handled.
It is important for PIDSes to reduce the number of false positives.

We category the false positives into three types in \S~\ref{sec:measurement-fpr}.
To reduce false positives, there are two possible ways.
The first way is to improve the generalization of PIDSes, for example, by adding potential variants of benign activities into the training data.
The other way is to identify potential false positives from the detection results.
For example, \orthrus uses a 2-class K-means to isolate outliers with the highest anomaly scores to further filter the potential false positives.
The former way requires sufficient prior knowledge of benign activities, which is difficult to obtain. Even worse, there will always be unknown benign activities. Therefore, we focus on the latter approach to reduce false positives.

Inspired by \orthrus, we design a process-level false positive detection algorithm based on the Louvain community detection algorithm. 
The algorithm divides processes with similar activities into the same community and considers frequently occurring processes as false positives.
On the one hand, this approach can reduce the FPRs. On the other hand, it can help analysts to investigate alerts more efficiently, as they only need to investigate one process in each community instead of all of them.
The details are provided in the Appendix~\ref{sec:appendix-fp}.
We take \nodlink as an example to reduce its false positives, as \nodlink is a process-level detection approach.
The results are shown in Fig.~\ref{fig:insight-fp_reduction}.
Fig.~\ref{fig:insight-fpr} shows that the false positive rates of \nodlink reduce from about 25\% to about 10\%.
Fig.~\ref{fig:insight-fp_num} shows the number of communities is about 1/3 of the number of false positives. An analyst only needs to investigate one process in each community instead of all false positives, which can greatly reduce the workload.

\noindent \textbf{Suggestion:} To reduce false positives, besides designing a more powerful PIDS, identifying potential false positives from detection results is also an effective way. What's more, detecting false positives while keeping attacks detected is a promising research direction.

\subsection{What problems may we overlook?}

\begin{figure}
    \centering
    \includegraphics[width=0.95\linewidth]{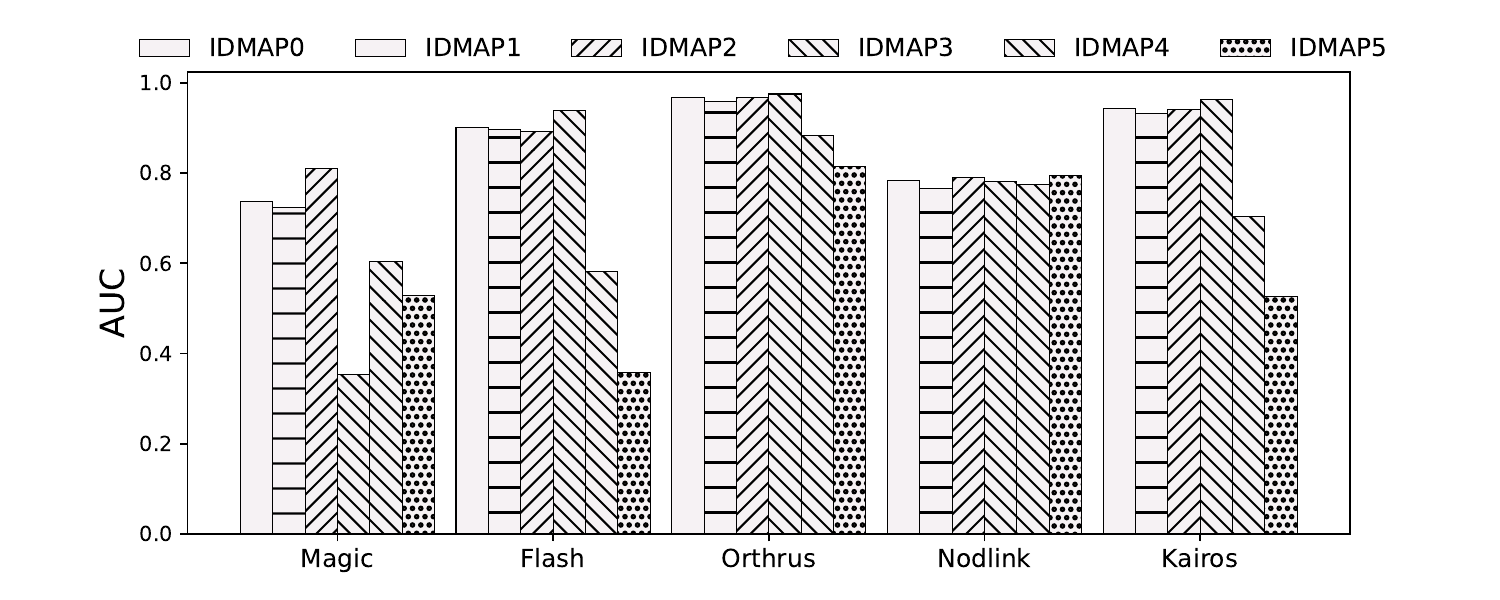}
    \caption{The AUCs of PIDSes under different uuid assignment strategies.}
    \label{fig:uuid-assign}
\end{figure}

\noindent \textbf{Insight 4: The way of assigning uuids to nodes affects detection.} 
In the DARPA datasets, each entity has already been assigned a uuid. However, such uuids may not be available in real scenarios. The uuid assignment strategy will impact the number of nodes, the graph structure, and the features of nodes, thus affecting the detection results.

% 溯源图中常见的节点类型为process、file、network socket。我们用文件路径为file节点生成uuid，而对于另外两种节点则有不同选择。对于process，我们使用进程对应的文件路径或PID+文件路径生成uuid。对于network socket，我们使用目标地址的IP或IP+端口号生成uuid。
The most common node types in provenance graphs are process, file, and network socket. 
We generate uuids based on file paths for file nodes. 
For the other two node types, there are various option. For process nodes, we use the file path of the relevant process or the combination of process ID (assigned by the operating system) and the file path to generate uuids. 
For network socket nodes, we use the destination IP address or a combination of the IP address and port number to generate uuids. Appendix~\ref{sec:appendix-idmap} provides more details about the uuid assignment strategies.

% 我们以Table~\ref{tab:eva-ideal-sec-host}作为基准，对不同策略进行了评估，结果如图~\ref{fig:uuid-assign}所示。不同方案对于uuid分配策略的敏感程度不同，例如，\orthrus和\nodlink的AUC在不同策略下变化较小，不超过10\%。而\magic的AUC变化较大，极差达到了0.8106 - 0.5293。
% 除了影响检测结果外，建图方式还影响节点的数量，最终影响检测时间。例如，当我们仅用目标地址IP生成uuid时，节点数目为4463.0。（额外考虑端口的话为24623.0）。此时各个方案的训练和检测时间大幅降低。
We use the dataset in Table~\ref{tab:eva-ideal-sec-host} for evaluation, and the results are shown in Fig.~\ref{fig:uuid-assign}.
Different PIDSes have varying sensitivities to uuid assignment strategies. \nodlink and \orthrus have small changes in AUC across different strategies, with variations within 10\%. In contrast, the AUCs of \magic change significantly, ranging from 52.93\% to 81.06\%.
Besides the detection performance, the uuid assignment strategy also affects the number of nodes, which eventually impact the time overhead. For example, when we generate uuids for network socket nodes using only the destination IP address, the number of nodes is 4,463, compared to 24,623 when port numbers are considered. This significantly reduces the training and detection time for all PIDSes.

% 建议：uuid的分配往往被忽略，然而实际其对检测结果和检测时间均有影响。我们建议未来的工作探寻更合理的uuid分配方式，以提升PIDSes的检测效果和效率。
\noindent \textbf{Suggestion:} How to assign uuids to nodes is often overlooked, yet it influences both detection performance and time overhead of PIDSes. We recommend to explore a more suitable uuid assignment strategy to reduce detection time while maintaining detection performance in future work.

\begin{table}[t]
\caption{TP and FP nodes and their corresponding entities in the real-case mining attack dataset.}
\label{tab:insight-detect-entity}
\centering
\resizebox{.80\linewidth}{!}{
\begin{tabular}{@{}ccccc@{}}
\toprule
\textbf{Method} & \textbf{TP} & \textbf{TP-entity} & \textbf{FP}  & \textbf{FP-entity} \\ \midrule
\magic & 14,808 & 125 & 112,846 & 9,854 \\
\flash & 113 & 113 & 8,368 & 8,368 \\
\orthrus & 0 & 0 & 6 & 6 \\
\kairos & 2,069 & 156 & 25,577 & 11,100 \\
\nodlink & 27,329 & 112 & 39,817 & 321 \\
\bottomrule
\end{tabular}
}
\end{table}

% PIDSes在检测后需要基于检测结果进行攻击调查。目前的方案大多为在根据alerts信息，在溯源图上进行前向后向追踪等方式定位攻击的源头与影响范围，从而帮助安全分析与攻击调查。然而，这种策略在fine-grained的检测上仍然存在一些问题需要解决。
\noindent \textbf{Insight 5: Filling the gap between fine-grained detection and attack investigation is needed.} 
PIDSes need to investigate attacks based on detection results. Existing approaches typically begin with the alerts and perform analysis such as forward and backward tracing to reconstruct the entire attack. 
However, when applying these approaches to fine-grained detection results, some problems still need to be addressed. A major problem we find is that fine-grained detection results do not fully align with the requirements of attack investigations, leading to high investigation overheads.

% 我们发现的一个最主要的问题在于fine-grained检测结果与调查需求并不完全匹配，使得调查开销较大。在表~\ref{tab:insight-detect-entity}中，我们统计了在真实挖矿攻击数据集中，各个PIDSes检测到的TP和FP节点数目及其对应的实体数目。可以看到，部分PIDSes的告警节点数远大于其对应的实体数目。如一个可执行文件被多次执行，被分配了多个PID，对应多个nodes。此时一个分析人员将分析多个溯源图（或节点较多的图），而实际上仅对应了同一个攻击。这意味着即便在误报较低的情况下，分析人员仍可能花费大量时间也仅能找到同一攻击，进一步加剧告警疲劳。
In Table~\ref{tab:insight-detect-entity}, we count the number of TP and FP nodes detected by each PIDS in the real-case mining attack dataset (as shown in Table~\ref{tab:eva-real-case-mining}), along with their corresponding entity counts. 
We observe that for some PIDSes, the number of alerted nodes is significantly higher than the number of corresponding entities. For instance, the malicious nodes detected by \nodlink is approximately 244 times the number of malicious entities. This occurs because a single executable file may be executed multiple times, resulting in multiple PIDs and multiple nodes in the provenance graph.
As a result, an analyst may need to analyze multiple graphs (or graphs containing many nodes) that correspond to the same attack. Even when the false positives are low, the analyst may spend a lot of time only to find the same attack, which further exacerbates alert fatigue.

% fine-grained检测结果虽然能精确到具体的节点，但其形式与调查需求并不完全匹配，导致调查开销较大。我们建议未来的工作探索如何将fine-grained检测结果与攻击调查需求更好地结合起来，以提升调查效率，减轻分析人员的负担。
\noindent \textbf{Suggestion:} Although fine-grained detection results can precisely identify specific malicious nodes, they may not fully align with the requirements of attack investigations, leading to high investigation overheads. We recommend future work to explore how to better integrate fine-grained detection results with attack investigation needs to improve investigation efficiency and reduce the burden on analysts.

\section{Discussion}
\label{sec:discussion}

% \noindent \textbf{The unification of data from multiple sources.}

\noindent \textbf{Adversarial manipulation and defences.}
In this paper, we do not discuss the adversarial attackers who may manipulate their activities to make them less detectable~\cite{mimic-attack}. 
On the one hand, this evaluation is already done by~\cite{survey-eval} on the DARPA-TC datasets. \cite{R-CAID} also proposes a method to defend against such attackers.
On the other hand, although we do not find or evaluate such attackers providing by~\cite{mimic-attack}, we find the attackers using other techniques to evade detection, such as the long-term information stealing attack.

\noindent \textbf{The limitations of our work.}
Although the data we used is more comprehensive than previous works, it still has some limitations. For example, we only use the data from one organization, and the data may not be representative of other organizations. What's more, due to the privacy and security concerns, we cannot share the data with the public, which may limit the reproducibility of our work. About the suggestions for the future work, we only provide some high-level directions, more specific and detailed evaluations and approaches are needed in the future work.
% \noindent \textbf{Rethinking rule-based detection approaches.}

% \noindent \textbf{Future work.}
% In the future work, we 

% More human-friendly explanations
% LLM for detection and investigatation 
% 日志缺失条件下的检测

\section{Related Work}
\label{sec:related-work}

In this section, we review related work about the survey of PIDSes. 
\cite{sok-pids} divides PIDSes into five layers and summarizes the corresponding techniques. This work mainly focuses on the log collection, protection and reduction. \cite{survey-provenance-prev,survey-provenance,anomaly-detection-survey} summarize the techniques used for attack detection. \cite{survey-industry} reviews PIDSes from the industrial perspective and provides lots of issues. \cite{survey-eval} conducts a comprehensive evaluation of SOTA anomaly-based PIDSes on DARPA-TC datasets.
As we know, our work is the first work to systematically evaluate the SOTA PIDSes on a real-world scenario.

% \noindent \textbf{Survey of provenance-based intrusion detection systems.}
% \cite{sok-pids, survey-industry, survey-provenance-prev, survey-provenance,survey-eval, anomaly-detection-survey}\cite{endpoint-security-survey,cloud-workload-protection-survey, ioc-measurement, fp-study}
% % survey-provenance-prev: summarize opportunities and challenges for endpoint intrusion detection systems
% % survey-provenance: the technical literature
% % survey-industry: industry solutions
% % cloud-workload-protection-survey : CWP
% % endpoint-security-survey : security terminal
% % ioc-measurement

% \noindent \textbf{Log collection and protection.}
% collection and reduction: \cite{Custos, linux-prov-capture, spade, hi-fi, ALASTOR, eaudit, palantir, LEONARD, nids-log-ebpf}
% protection: \cite{logging-system, hardlog, KennyLoggings, logging-system}

% \noindent \textbf{Attack detection.}
% rule-based: \cite{SLEUTH, SAQL, MORSE, HOLMES, Poirot, RapSheet, p-gaussian, CAPTAIN}
% anomaly-based: \cite{Deeplog, UNICORN, SIGL, StreamSpot, Log2vec, provtalk, PROVDETECTOR, SHADEWATCHER, THREATRACE, PROGRAPHER, NODLINK, DISTDET, KAIROS, R-CAID, FLASH, MAGIC, orthrus,slot, STGAN, llm-detection}

% \noindent \textbf{Attack investigation.}
% \cite{OmegaLog, NODOZE, ATLAS, MPI, WATSON, DEPIMPACT, Provg-searcher, ALchemist, uiscope, mci, PRIOTRACKER, DEEPCOM, TREC, Contexts} 
% mimic attack: \cite{mimic-attack, R-CAID, PROVNINJA}

\section{Conclusion}
\label{sec:conclusion}

In this paper, we conduct systematic evaluation and analysis of PIDSes in the industrial scenarios. We first conclude three main new characteristics in industry: heterogeneous multi-source inputs, more powerful attackers, and increasing benign activity complexity.
Then, we select five state-of-the-art PIDSes and build several datasets to evaluate the performance of these PIDSes. The evaluation results reveal challenges for existing PIDSes, including poor portability across different hosts and platforms, low detection performance against real-world attacks, and high false positive rates with ever-changing benign activities. 
Based on the evaluation results and our industrial practices, we provide several insights that aim to solve the above problems and improve PIDSes.
We hope our work can inspire other researchers to design more effective PIDSes and deploy them in real-world industrial scenarios.

\balance{}
\bibliographystyle{IEEEtran}
\bibliography{ref}

\appendix
\subsection{Distances between nodes via the type information}
\label{sec:appendix-distance}

We show how we calculate the distance between two nodes via the type information for \magic. For each node $v$, we start at $v$ and use a depth first search to record the types of all neiboring nodes with edges, which is defined as:
$$
Info(v): \{(type(u), type(u\rightarrow w), type(w)), u,w \in \mathcal{N}_k(v)\},
$$
where $type()$ is the type of a node or an edge, and $\mathcal{N}_k(v)$ is the set of nodes that are $k$-hop neighbors of $v$. Then, we represent $v$ as a vector $x_v$ by counting the number of occurrences of each type in $Info(v)$, as: 
$$
x_v = [c_0,...,c_s],c_i \in \mathbb{N},
$$
where $s$ the total number of type tuples $(type(u), type(u\rightarrow w), type(w))$ in the graph, and $c_i$ is the number of occurrences of the $i$-th type tuple in $Info(v)$. Finally, we calculate the distance between two nodes $v$ and $v'$ using euclidean distance between their vector representations, as:
$$
d(v,v') = \sqrt{\sum_{i=0}^s (c_i - c_i')^2}.
$$

In the evaluation in \S~\ref{sec:insight}, we set $k=2$ as \magic only considers 2-hop neighbors for embedding generation. What's more, to make sure the efficiency of the evaluation, we consider at most 100 neighbors for each node.

\subsection{False positive detection}
\label{sec:appendix-fp}

We show the details of the false positive detection approach used in \S~\ref{sec:insight}. The approach contains two main steps: (1) We represent each process as a vector with its behavior sequence. (2) We use the Louvain community detection algorithm to cluster the processes into different communities, and we consider the processes in the large-size communities as false positives.

A process $p$ always has plenty of behavior, which can be represented as:
$$
p:\{(action,\ object,\ timestamp)\},
$$
where $action$ contains operations such as file read/write, and $object$ can be a process, a file, or a network connection.
We refer to term frequency-inverse document frequency (TF-IDF) to represent the process $p$ as a vector $x_p$, which is defined as:
$$
x_p = [freq(object_i)*\log(\frac{N_{p}}{n_i}),...],
$$
where $object_i$ is the $i$-th object in the graph, $freq(object_i)$ is the times that $p$ interacts with $object_i$, $N_p$ is the total number of processes, and $n_i$ is the number of processes that interact with $object_i$.
We use the euclidean distance to calculate the similarity between two processes. Then, we apply the Louvain community detection algorithm to cluster the processes into different communities. Finally, we deem the processes in the communities whose size is larger than 20 as false positive processes.

\subsection{UUID assignment to nodes}
\label{sec:appendix-idmap}

\begin{table}[t]
\centering
\caption{The fields used for UUID generation for different types of nodes.}
\label{tab:idmap}
\begin{tabular}{cc}
\hline
\textbf{Node Type} & \textbf{Fields for UUID Generation} \\ \hline
process & PID, file path \\
file & file path \\
\multirow{2}{*}{network} & domain $>$ url $>$ source IP, source port, \\
& destination IP, destination port \\
registry Key & file path \\
script & script content \\
\hline
\end{tabular}
\end{table}

We extract different fields for different types of nodes and use a hash function to generate UUID for each node. The details of the fields used for different types of nodes are shown in Table~\ref{tab:idmap}.
For network nodes, we use the domain if it exists, then the url, and finally the source and destination IP and port. We choose this order because the domain and url are more stable than the IP and port.
In Fig.~\ref{fig:uuid-assign}, we change the fileds used for the process nodes and network nodes. Our changes include: (1) removing the PID for process nodes, (2) removing the domain and url for network nodes, (3) only using the source IP and destination IP for network nodes, and (4) only using the destination IP for network nodes.
IDMAP1 contains the change (2). IDMAP2 contains the change (3). IDMAP3 contains the change (1). IDMAP4 contains the change (4). IDMAP5 both contains the change (1) and (4).


% \begin{IEEEbiography}[{\includegraphics[width=1in,height=1.25in,clip,keepaspectratio]{figures/demo-llm.pdf}}]{Michael Shell}
% \end{IEEEbiography}

% \vfill

\end{document}